\documentclass[prx,twocolumn]{revtex4-1}
\usepackage{amsmath,amsfonts,graphicx,enumerate,color}
\usepackage{bm}
\usepackage{tabularx}
\newcommand{\pa}{\partial}

\newcommand{\tel}{\tau^{el}}

\newcolumntype{Y}{>{\centering\arraybackslash}X}
\begin{document}
\title{Velocity-strengthening friction significantly affects\\ interfacial dynamics, strength and dissipation}

\author{Yohai Bar-Sinai$^1$, Robert Spatschek$^2$, Efim A. Brener$^{1,3}$ and Eran Bouchbinder$^1$}
\affiliation{$^1$Chemical Physics Department, Weizmann Institute of Science, Rehovot 76100, Israel
\\$^2$Max-Planck-Institut f\"ur Eisenforschung GmbH, D-40237 D\"usseldorf, Germany
\\$^3$Peter Gr\"unberg Institut, Forschungszentrum J\"ulich, D-52425 J\"ulich, Germany}

\begin{abstract}
Frictional interfaces are abundant in natural and man-made systems and their dynamics still pose challenges of fundamental and technological importance. A recent extensive compilation of multiple-source experimental data has revealed that velocity-strengthening friction, where the steady-state frictional resistance increases with sliding velocity over some range, is a generic feature of such interfaces. Moreover, velocity-strengthening friction has very recently been linked to slow laboratory earthquakes and stick-slip motion. Here we elucidate the importance of velocity-strengthening friction by theoretically studying three variants of a realistic rate-and-state friction model. All variants feature identical logarithmic velocity-weakening friction at small sliding velocities, but differ in their higher velocity behaviors. By quantifying energy partition (e.g. radiation and dissipation), the selection of interfacial rupture fronts and rupture arrest, we show that the presence or absence of velocity-strengthening friction can significantly affect the global interfacial resistance and the total energy released during frictional instabilities (``event magnitude''). Furthermore, we show that different forms of velocity-strengthening friction (e.g. logarithmic vs. linear) may result in events of similar magnitude, yet with dramatically different dissipation and radiation rates. This happens because the events are mediated by interfacial rupture fronts with vastly different propagation velocities, where stronger velocity-strengthening friction promotes slower rupture. These theoretical results may have significant implications on our understanding of frictional dynamics.
\end{abstract}

\maketitle

\section{Introduction}
\label{intro}

Frictional interfaces are abundant in biological (e.g. adherent cells and cell locomotion), engineering (e.g. micro-electro-mechanical devices) and geophysical (e.g. Earthquake faults) systems around us, and are of fundamental and practical importance. Consequently, understanding the dynamics of dry frictional interfaces has been the focus of intense scientific activity in the last few decades
\cite{Dieterich1978, Ruina1983, Marone1998, Persson2000, Scholz2002, Baumberger2006}. It has been established that under
steady-state sliding conditions, the frictional resistance features a non-trivial velocity dependence, and that this
dependence has dramatic consequences on the dynamic response of frictional interfaces \cite{Rice1983, Gu1984, Heslot1994,
Baumberger1999, Rice2001, Kaproth2013}. Specifically, it has been shown that for a broad range of materials
friction is velocity-weakening -- that is, the steady frictional resistance is a decreasing function of the sliding
velocity -- at least in the regime of low velocities, up to a few hundreds of microns per second. This feature favors various instabilities and stick-slip motion \cite{Marone1990, Scholz2002, Rubin2005, Liu2005}.

A very recent compilation of a large set of experimental data for a broad range of materials, however, has revealed that for higher slip velocities, friction generically becomes velocity-strengthening over some range of slip velocities \cite{Bar-Sinai2013jgr}. The existence of velocity-strengthening behavior might have significant effects on various aspects of frictional dynamics. In particular, recent laboratory experiments on fault-zone materials have documented slow slip interfacial events -- an intensely debated issue -- and have linked it to a crossover in the frictional response, from velocity-weakening to velocity-strengthening friction, with increasing slip velocity \citep{Kaproth2013}. While the possible implications of the existence of velocity-strengthening friction have been rather sporadically discussed in the literature \cite{Kaproth2013, Weeks1993, Kato2003, Shibazaki2003, Bouchbinder2011, BarSinai2012, Hawthorne2013, Bar-Sinai2013pre, Hawthorne2013, Rice2001, BarSinai2012, Marone1991, 02BM, 05BMM}, to the best of our knowledge a comprehensive and systematic theoretical exploration of these important issues is currently missing.

As a first step in closing this gap, we study here the effect of velocity-strengthening friction on spatiotemporal interfacial dynamics, energy dissipation and radiation, and the global interfacial strength, with a special focus on the nucleation, propagation and arrest of rupture fronts. We explore three variants of a realistic rate-and-state friction law, one which is purely velocity-weakening, one which crosses over at higher velocities to logarithmic velocity-strengthening friction, and one which crosses over to linear velocity-strengthening friction.

We show that the presence or absence of velocity-strengthening friction at relatively high slip velocities can significantly affect the global interfacial resistance (strength) and the energy released during frictional instabilities (``event magnitude''), even under quasi-static loading conditions. Different forms of velocity-strengthening friction, in our case logarithmic and linear, give rise to events of similar magnitude, yet with dramatically different dissipation and radiation rates. The difference stems from the broad range of the underlying rupture propagation velocities, where stronger velocity-strengthening friction promotes slower rupture, possibly orders of magnitude slower than elastic wave-speeds. This result is directly related to the recent experimental observations of \citep{Kaproth2013}. All in all, our results show that velocity-strengthening friction should be properly quantified and incorporated into friction theory as it appears to affect many basic properties of spatially extended frictional interfaces.

\section{Theoretical considerations}
\label{models}

The ideas to be presented below have been originally influenced by the works in \cite{02BM, 05BMM, Braun2009} and then further developed in \cite{Bouchbinder2011, BarSinai2012, Bar-Sinai2013pre, Bar-Sinai2013jgr}.
The rate-and-state friction model we study has been introduced recently in \cite{BarSinai2012, Bar-Sinai2013pre}, and is
reviewed here briefly. We start by considering a multi-contact interface and write the ratio $A$ of the real contact area (the area of all contact asperities) to the nominal one, in terms of a state parameter $\phi$ (of time dimensions) as
\begin{equation}
 A(\phi)=\frac{\sigma}{\sigma_{\hbox{\tiny H}}}\left[1+b\log\left(1+\frac{\phi}{\phi^*}\right)\right] \ ,
 \label{eq:A}
\end{equation}
where $\sigma$ is the normal (compressive) stress at the interface, $\sigma_{\hbox{\tiny H}}$ is the material hardness,
$b$ is a dimensionless material parameter of order $10^{-2}$, and $\phi^*$ is a short time cutoff \cite{Nakatani2006,
Ben-David2010-ageing, Putelat2011, Bar-Sinai2013jgr}. $\phi$ is usually interpreted as the interface's effective age,
and its
evolution is given by
\begin{equation}
 \pa_t\phi=1-\frac{\phi\,v}{D}g(v)  \ ,
 \label{eq:phidot}
\end{equation}
where $v$ is the local interfacial slip velocity and $D$ is a lengthscale related to the contact asperities geometry. $g(v)\!=\!\sqrt{1+(v_0/v)^2}$, with an extremely small $v_0\!=\!1$nm/s, is a regularization function that plays no important role, and is actually omitted in all of the analytic results that follow. As we focus here on unidirectional motion, we do not distinguish between $v$ and $|v|$.

The frictional stress $\tau$ is written as a sum of an elastic contribution, $\tau^{el}$, and a viscous contribution
$\tau^{vis}$,
\begin{equation}
 \tau=\tau^{el}+\tau^{vis} \ .
 \label{eq:KV}
\end{equation}
The viscous contribution takes the form $\tau^{vis}\!=\!A(\phi)\,w(v)$, where at least at low
velocities, the rheological part $w(v)$ corresponds to a stress-biased thermally-activated process \cite{Rice2001, Putelat2011, Bar-Sinai2013jgr}
\begin{equation}
 w\left(v\right)=\frac{k_B T}{\Omega}\log\left(1+\frac{v}{v^*}\right)
\label{eq:tauvis} \ .
\end{equation}
Here, $k_B$ is Boltzmann's constant, $T$ is the absolute temperature, $\Omega$ is an activation volume and $v^*$ is a
velocity scale, related to a microscopic attempt rate. A higher velocity variant of Eq. (\ref{eq:tauvis}) will be discussed below.

The elastic stress follows the evolution equation
\begin{equation}
 \pa_t\tau^{el}= \frac{G_0 }{h}A\,v-\tel \frac{v}{D}g(v)\ ,
 \label{eq:tauel}
\end{equation}
where $G_0$ is the interfacial shear modulus and $h$ is the effective height of the interface.

Equations \eqref{eq:A}-\eqref{eq:tauel} describe the first variant of the friction model we study below. We begin by
describing its behavior under steady sliding at a velocity $v_d$. The steady solution of Eq. \eqref{eq:phidot} is
$\phi_{ss}(v)\!=\!D/v$, from which it follows that the contact area is a logarithmically decreasing function of $v$
\cite{Dieterich1978,Tullis1986,Nagata2008}. The fixed point of Eq. \eqref{eq:tauel} reads $\tau^{el}_{ss}(v)\!=\!G_0 D A\big(\phi_{ss}(v)\big)/h$, and hence
the overall frictional resistance is given by
\begin{equation}
f_{ss} \equiv \frac{\tau_{ss}(v)}{\sigma} \simeq f_0+\alpha
\log\left(1+\frac{v}{v^*}\right)+\beta\log\left(1+\frac{D}{v \phi^*}\right)\ ,\label{eq:RSF}
\end{equation}
where a higher order logarithmic term was omitted and the following definitions were used
\begin{equation}
\alpha \equiv \frac{k_B T}{\sigma_{\hbox{\tiny H}} \Omega} \ , \quad
\beta \equiv \frac{G_0 D b}{h\,\sigma_{\hbox{\tiny H}}\Omega} \ , \quad
f_0\equiv\frac{\beta}{b} \ .
\label{eq:rsf^defs}
\end{equation}
In the low velocity regime, i.e. $v\!\ll\!D/\phi^*$, $f_{ss}$ is a logarithmic function of $v$, with $\pa
f_{ss}/\pa\log v \approx \alpha-\beta$. Therefore, if $\alpha\!<\!\beta$ (which is quite generically the case), friction is
logarithmic velocity-weakening.

\begin{figure}
 \centering
 \includegraphics[width=\columnwidth]{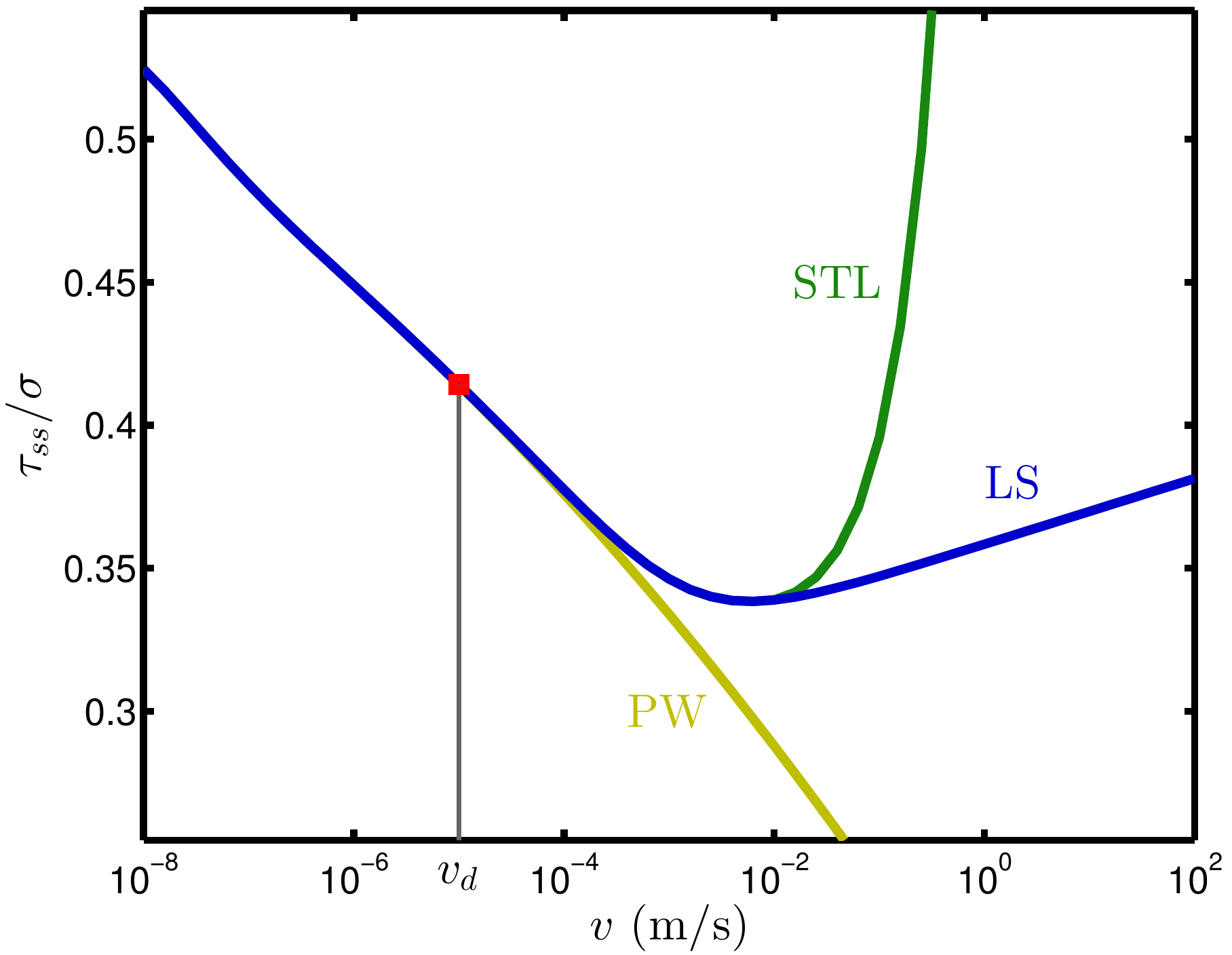}
 \caption{The steady sliding friction coefficient $\tau_{ss}/\sigma$ vs. the slip velocity $v$ for the three model variants (different colors, also marked by labels). Note that all of the curves coincide at low velocities and that the driving velocity $v_d$ is marked.}
 \label{fig:ss}
\end{figure}

Physically, friction is velocity-weakening because the real contact area is a decreasing function of the
sliding velocity, and its velocity dependence is stronger than the rheological dependence of $\tau^{vis}$.
However, as discussed at length in \cite{BarSinai2012, Bar-Sinai2013jgr, Bar-Sinai2013pre}, when $v \!\gtrsim\! D/\phi^*$
the contact area saturates, and friction becomes logarithmically velocity-strengthening. We term this model the
\textit{logarithmic velocity-strengthening} (LS) friction model. The resulting steady-state friction curve is shown in
Fig. \ref{fig:ss}.
In case the contact area continues to decrease indefinitely with increasing $v$, friction remains velocity-weakening for arbitrarily high
velocities. This is formally achieved by removing the ``1'' in the argument of the logarithm in Eq. \eqref{eq:A}, that
is, replacing Eq. \eqref{eq:A} by
\begin{equation}
 A(\phi)=\frac{\sigma}{\sigma_{\hbox{\tiny H}}}\left[1+b\log\left(\frac{\phi}{\phi^*}\right)\right] \ .
 \label{eq:A_wrong}
\end{equation}
Although this is somewhat unphysical, this choice was widely used in the literature \cite{Marone1998, Rice2001,
Nakatani2001, Scholz2002, Ampuero2008}, and we term it the \textit{pure velocity-weakening} (PW) friction model. The resulting
steady-state friction curve is also shown in Fig. \ref{fig:ss}.

A third variant of the model is obtained by modifying the rheological function $w(v)$, cf. Eq. \eqref{eq:tauvis}. As discussed
extensively in \cite{Bar-Sinai2013jgr}, and to some extent in \cite{Baumberger2006}, the simple picture of a single
barrier, linearly biased, thermally-activated process is expected to break down when asperity-level stresses become sufficiently large. When
this happens, a different dissipation mechanism is expected to dominate friction. While at the moment there is no
general quantitative theory for the velocity dependence of friction in this regime, it is not expected to be logarithmic, but rather to exhibit a significantly stronger dependence on the slip velocity.
In this work, we consider a simple model in which the logarithmic
dependence crosses over continuously (but not smoothly) to a linear viscous rheology. Explicitly, we replace Eq. \eqref{eq:tauvis} by
\begin{equation}
 w\left(v\right)=\begin{cases}
                 \frac{k_B T}{\Omega}\log\left(1+\frac{v}{v^*}\right) & v\le v_c\\
                 \frac{k_B T}{\Omega}\left[\log\left(1+\frac{v_c}{v^*}\right)+m\left(\frac{v}{v_c}-1\right)\right] & v>
v_c\\
                 \end{cases}\ ,
\label{eq:tau_LTS}
\end{equation}
where $m$ is a dimensionless parameter. We term this model the
\textit{stronger-than-logarithmic} (STL) velocity-strengthening model. The resulting steady-state friction curve is shown in
Fig. \ref{fig:ss}.

We stress that all three variants coincide in the low velocity regime, where they
feature logarithmic velocity-weakening friction. At higher slip velocities, the LS variant, which is described by Eqs. \eqref{eq:A}-\eqref{eq:tauel},
features a crossover to logarithmic velocity-strengthening friction. The PW variant does not feature any strengthening at all (i.e. it remains velocity-weakening), and is
obtained from the LS model by using Eq. \eqref{eq:A_wrong} instead of \eqref{eq:A}. The STL variant features linear
velocity-strengthening friction, and is obtained from the LS model by using Eq. \eqref{eq:tau_LTS} instead of Eq. \eqref{eq:tauvis}.

\begin{figure}[here]
 \centering
 \includegraphics[width=\columnwidth]{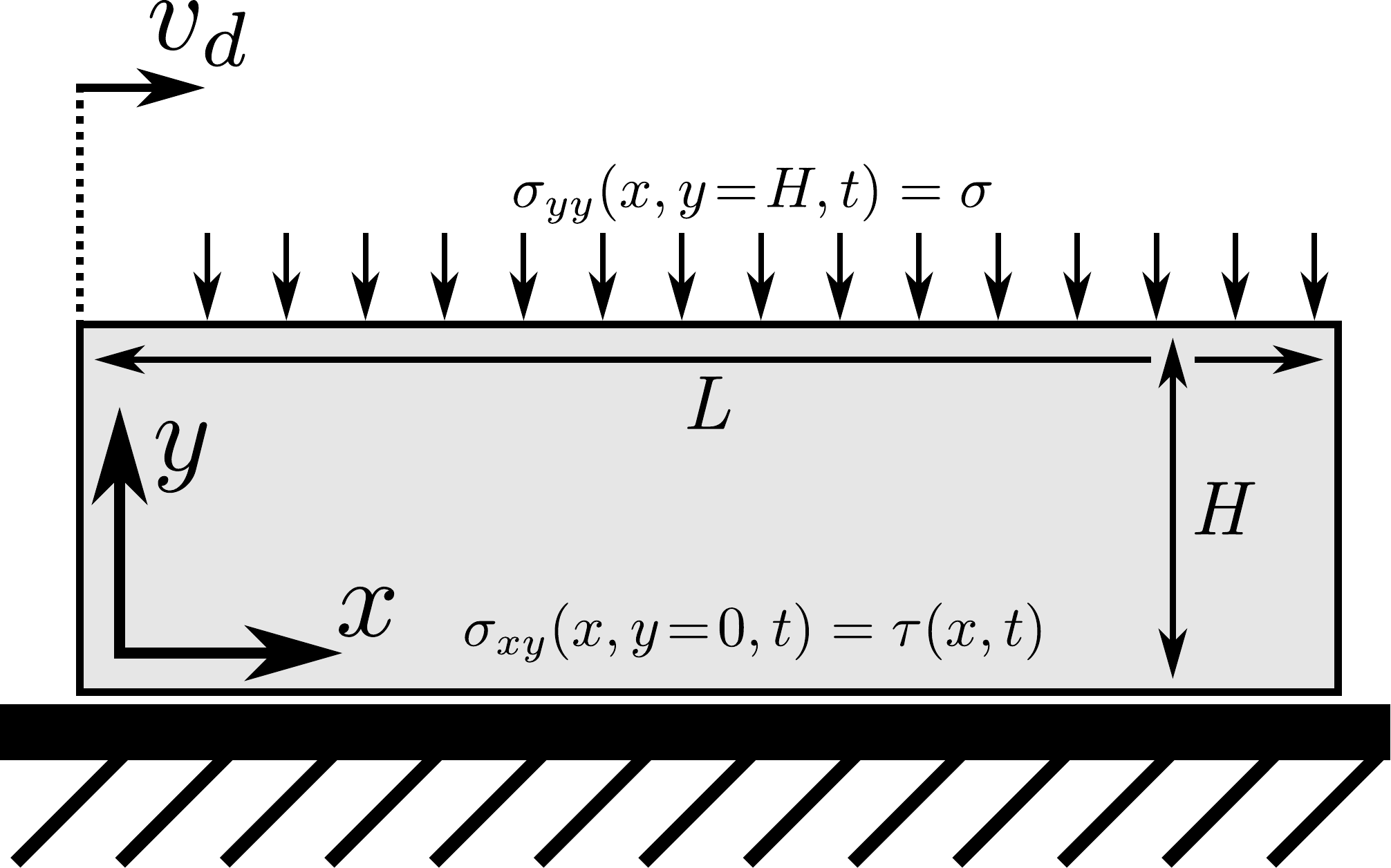}
 \caption{A sketch of the spatially-extended frictional system. An elastic block, which is in frictional contact with a rigid substrate, is loaded by a space- and time-independent normal stress $\sigma_{yy}(x,y\!=\!H,t)\!=\!\sigma$ ($H$ is the block's height) and driven by a velocity $v_d$ at its trailing edge ($x\!=\!0$). The leading edge is at $x\!=\!L$. The shear stress at the interface, $\sigma_{xy}(x,y\!=\!0,t)$, equals to the frictional stress $\tau(x,t)$.}
 \label{fig:Sketch}
\end{figure}

In order to investigate the implications of the different constitutive laws on frictional dynamics, we need to consider a spatially-extended
interface under inhomogeneous sliding conditions. To this end, we consider a long elastic block of height $H$ (in the $y$-direction) and length $L\!\gg\!H$ (in the $x$-direction), in frictional contact (at $y\!=\!0$) with a rigid substrate (i.e. no deformation of the substrate is considered), see Fig. \ref{fig:Sketch}. The trailing edge of the elastic block (at $x\!=\!0$) is moved at a constant velocity $v_d$ in the positive $x$-direction, while the leading edge (at $x\!=\!L$) is stress-free. The block is driven quasi-statically with $v_d\!=\!10\mu$m/s, which is representative of typical laboratory experiments \cite{Rubinstein2007,Ben-David2010} and generically belongs to the steady-state velocity-weakening friction branch (cf. Fig. \ref{fig:ss}). The upper edge of the elastic block (at $y\!=\!H$) experiences a constant normal stress $\sigma$, $\sigma_{yy}(x, y\!=\!H,t)\!=\!\sigma$, but no shear stress, i.e. $\sigma_{xy}(x,y\!=\!H,t)\!=\!0$.

We focus on plane-strain deformation conditions and furthermore assume that $H$ is smaller than the smallest lengthscale $\ell$ characterizing the spatial variation of various fields in the $x$-direction. Under the stated conditions, the momentum balance equation
\begin{equation}
\label{eq:momentum}
\rho\pa_{tt} u_i\!=\!\pa_j\sigma_{ij} \ ,
\end{equation}
where $u_i$ and $\sigma_{ij}$ ($i,j\!=\!x,y$) are the components of the displacement vector and Cauchy's stress tensor, respectively, and $\rho$ is the mass density, reduces to (see \cite{Bar-Sinai2013pre} for derivation)
\begin{eqnarray}
\label{eq:EOM1}
&\rho H\pa_{tt}u = \bar{G}H\pa_{xx}u-\tau \ ,\\
&\sigma_{yy}(x,y,t) = \sigma \ ,
\label{eq:EOM2}
\end{eqnarray}
where the plane-strain Hooke's law was used. Here
\begin{equation}
u(x,t) \equiv \frac{1}{H}\int_0^H \!\!u_x(x,y,t)\,dy \ ,
\end{equation}
$\bar{G}\!=\!\frac{2G}{1-\nu}$ (where $G$ is the shear modulus of the bulk and $\nu$ is Poisson's ratio) and the shear stress at $y\!=\!0$ simply equals the frictional stress, $\sigma_{xy}(x,y\!=\!0,t)\!=\!\tau(x,t)$. Note also that $v(x,t)\!=\!\pa_t u(x,t)$. Corrections to Eqs. (\ref{eq:EOM1})-(\ref{eq:EOM2}) appear only to order $\left(H/\ell\right)^2$, a situation reminiscent of the shallow water approximation in fluid mechanics. Finally, note that the lateral force required to maintain the velocity boundary condition at the trailing edge, $u(x\!=\!0,t)\!=\!v_d\, t$, reads
\begin{equation}
f_d(t)=-\left.\bar{G}H\pa_x{u(x,t)}\right|_{x=0}\ ,
\label{eq:fd}
\end{equation}
and the traction-free boundary condition at the leading edge implies $\pa_x u(x\!=\!L,t)\!=\!0$.

Equation \eqref{eq:EOM1}, with the stated boundary conditions and with $\tau(x,t)$ corresponding to one of the three friction laws described above, has been solved numerically using a commercial differential equations solver. The model parameters for polymethyl-methacrylate(PMMA), a
polymeric glass that is widely used in laboratory experiments \cite{Berthoud1998a, Bureau2003, Rubinstein2007,
Ben-David2010}, were extracted from a large set of experimental data. The parameters are listed in Table
\ref{tab:params}, and the procedure for obtaining them is described in \cite{Bar-Sinai2013pre}. The initial conditions are
$u(x,t)\!=\!0$, $v(x,t)\!=\!0$, $\tau(x,t)\!=\!\tau^{el}(x,t)\!=\!0$, and $\phi(x,t)\!=\!1$s, the latter is typical of laboratory scale experiments. The results
presented here are largely insensitive to the choice of the initial value of $\phi$.

\begin{table}[here]
 \centering
 \bgroup
\def\arraystretch{1.5}
 \begin{tabular}{|c|c||c|c|c}
  \hline
 $\bar{G}$ & 9.3 GPa &  $\sigma$ & 1 MPa \\ \hline
 $G_0/h$& 300 MPa/$\mu$m &  $b$ & 0.075\\ \hline
 $\rho$ & 1,200 Kg/m$^3$ & $D$ & 0.5 $\mu$m  \\ \hline
 $v^*$ & 0.1 $\mu$m/s & $D/\phi^*$ & 1.5 mm/s  \\ \hline
 $\frac{k_B T}{\Omega}$ & 27 MPa & $\sigma_{\hbox{\tiny H}}$ & 540 MPa\\ \hline
 $m$ & 25 & $v_c$ & 7.5 mm/s\\ \hline
 \end{tabular}
 \egroup
  \caption{Material parameters for PMMA\footnote{Except for $m$ and $v_c$, which have not been yet directly measured for this material. For such measurements in other materials, see Fig. 1 in \cite{Bar-Sinai2013jgr}.}.}
 \label{tab:params}
\end{table}

\section{Results}
\subsection{Global frictional resistance}

\begin{figure}
 \centering
 \includegraphics[width=\columnwidth]{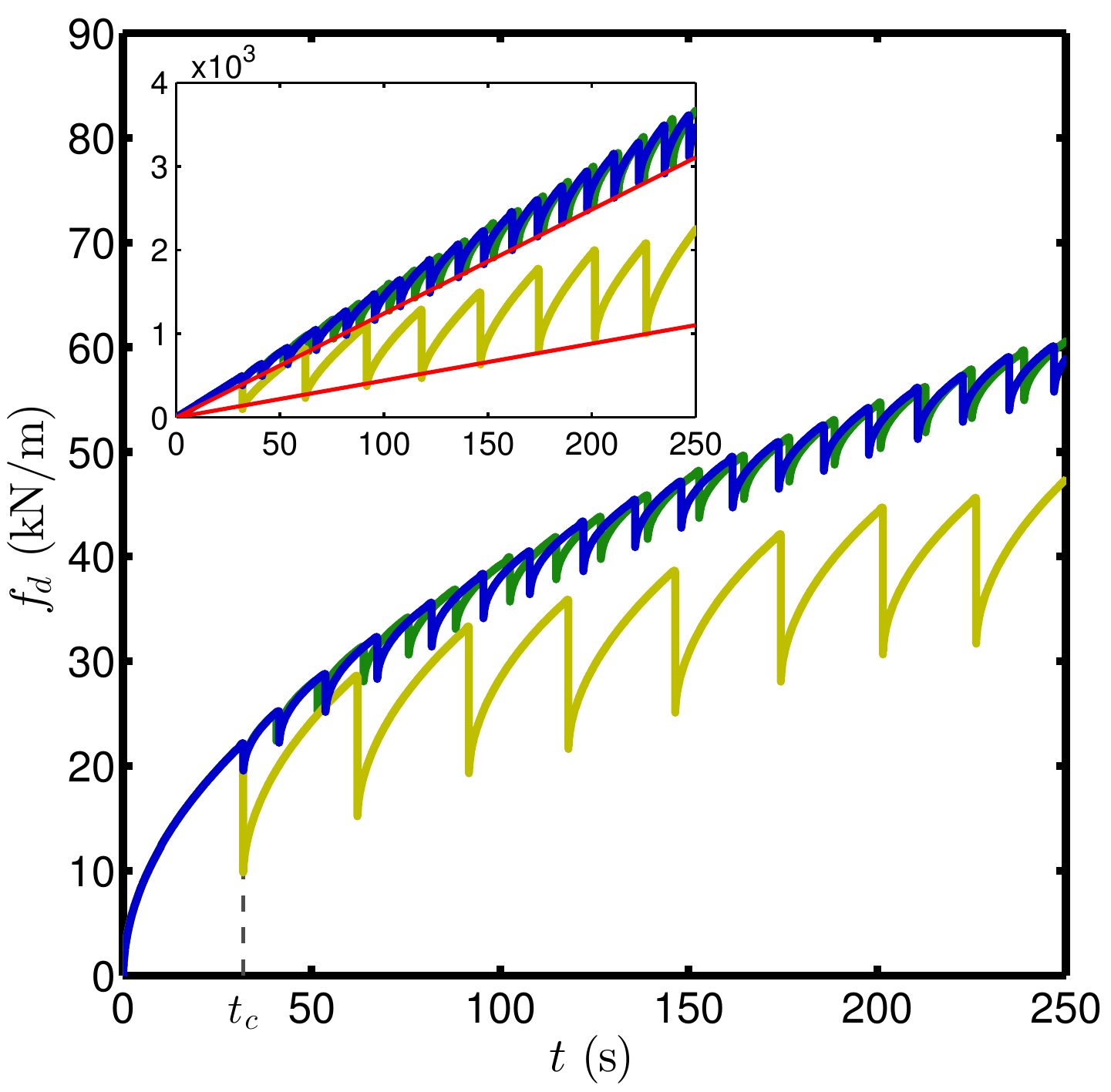}
 \caption{The loading force $f_d$ vs. time for the three models (color code as in Fig. \ref{fig:ss}). It is seen that
  all of the curves coincide for short times, and then begin to diverge. The LS and STL models maintain the same
``envelope'', while the PW  model features more pronounced stress drops, larger inter-event time and a lower overall resistance. (inset) The same data as in the main panel, but this time $f_d^2$ is plotted vs. time. The red lines are linear fits to the values of $f_d^2$ at the rupture arrest times $t_a$ (i.e. $f_d^2$ right after the force drops), cf. the prediction in Eq. \eqref{eq:fd_xtip}.}
 \label{fig:fd}
\end{figure}

We begin by studying the macroscopic response of the system. Figure \ref{fig:fd} shows the total frictional force
exerted by the loading machine as a function of time, $f_d(t)$. It is seen that the friction force increases gradually until it
experiences an abrupt drop, followed by repeated cycles of gradual increases and abrupt drops, typical of frictional systems \cite{Rubinstein2007, Tromborg2011, Otsuki2013}. The drops in the friction force, which appear as vertical lines in this figure, occur when sliding becomes unstable, and involve nucleation and propagation of rupture fronts, as will be discussed below.

Before the first drop, the friction force corresponding to the three variants is identical, as can be expected because the
dynamics in this regime are slow and governed by the loading velocity $v_d$. In this range of velocities, the three
variants coincide and consequently the first drop occurs almost exactly at the same point in time for all of
the variants, suggesting that the instability mechanism is insensitive to the high velocity behavior (as
predicted in \cite{Bar-Sinai2013pre}). However, since the instabilities are accompanied by much larger
velocities, the high velocity behavior of the friction law becomes important.

Figure \ref{fig:fd} demonstrates that while the LS and STL models give rise to almost identical force profiles, the PW
model results in significantly larger force drops, and a lower overall interfacial resistance. This suggests, and will be further substantiated in
what follows, that while the total energy dissipated during these drops is similar in the LS and STL models, the energy
dissipated in the PW model is significantly larger. Other features of the global friction curves shown in Fig. \ref{fig:fd}, such as the lower envelope of $f_d(t)$ (corresponding to the values of $f_d(t)$ after each drop), will be discussed and explained theoretically below.

\subsection{Spatiotemporal interfacial dynamics}

\begin{figure}
\centering
\includegraphics[width=\columnwidth]{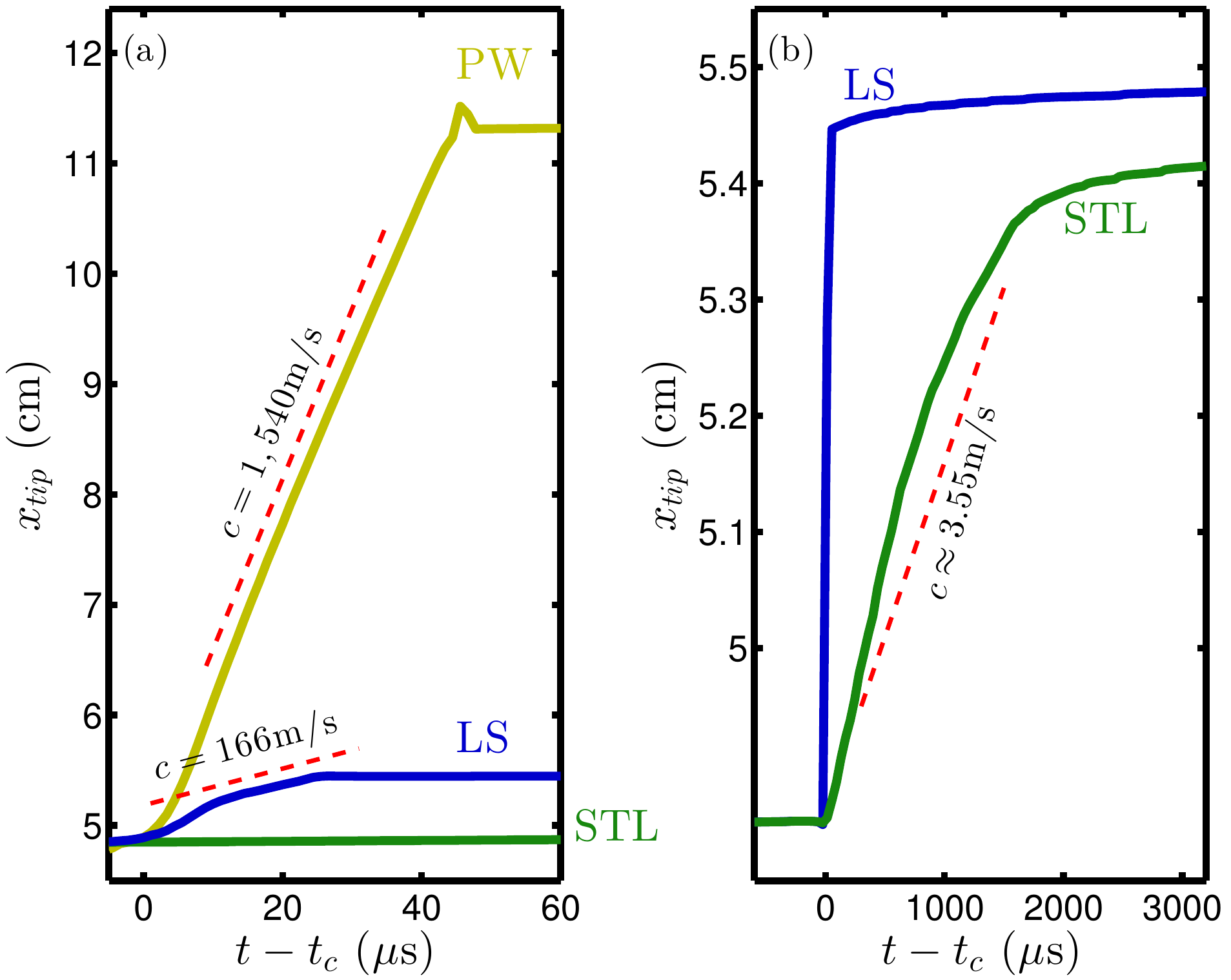}
\caption{Propagation of rupture fronts in the first event for (a) All three models and (b) the LS and LTS models.
$x_{tip}$ is the spatial location of the front tip, cf. Fig. \ref{fig:before_after}. $t_c$ is the time where the front starts to propagate, cf. Fig. \ref{fig:fd}. Note also the vast difference in
timescales between the panels. The wave speed in this system is $\sqrt{\bar{G}/\rho}\approx 2700$ m/s.}
\label{fig:fronts}
\end{figure}

In order to understand the origin of these differences, one must examine the complex spatiotemporal dynamics that
give rise to the ``force drop events'', which are described at length in \cite{Bar-Sinai2013pre}. As stated above,
the instabilities result in the nucleation, propagation and arrest of rupture fronts, a scenario reported by many experimental, numerical and analytical works \cite{Ohnaka2000, Rubinstein2004, Braun2009, Ben-David2010-fronts, Kammer2012, Latour2013}. Most of the remainder of this paper
will be focused on the first event, which is marked in Fig. \ref{fig:fd} by $t_c$. The rationale for focussing on the first event (rather than some later event) is that it ensures that the state of the interface is the same for all three model variants at the onset of instability (with no history effects), cleanly isolating the effects of the existence and form of the velocity-strengthening branch. Having said that, we note that it is clear from Fig. \ref{fig:fd} that the differences between the three variants persist to {\em any} event. Furthermore, multiple-event properties will be explicitly discussed in relation to Eqs. \eqref{eq:approx_fd}-\eqref{eq:fd_xtip} and the inset of Fig. \ref{fig:fd}.

Figure \ref{fig:fronts} shows the propagation and arrest of rupture fronts during the first event. First, we note the vast difference in the timescales involved:
while rupture fronts in the LS and PW models arrest after a few 10$\mu$s, in the STL model they last for a few ms. It is observed, however, that while the penetration depth of
the front into the interface in the LS and STL models is comparable, for the PW model it is an order of magnitude larger. Furthermore, the rupture propagation velocity in the LS model is an order of magnitude smaller than in the PW model (the latter is of the order of the elastic wave-speed), and the propagation velocity in the STL model is yet two orders of magnitude smaller.

Both the LS and STL models give rise to rupture fronts that are much slower than the elastic wave-speed.
These remarkably low rupture propagation velocities, three orders of magnitude slower than the elastic wave-speed in the STL model, might
be related to the important, and rather intensely debated, issue of slow rupture \cite{Peng2010, Kato2012, Hawthorne2013, Ikari2013, Tromborg2014}. Our
calculations suggest that the emergence of slow rupture might be directly related to the existence and form of
velocity-strengthening friction. This is in accord with recent laboratory experiments on fault-zone materials, which
documented slow slip events together with a clear crossover from velocity-weakening to velocity-strengthening
friction with increasing slip velocity \citep{Kaproth2013}.

\begin{figure}
\centering
\includegraphics[width=0.95\columnwidth]{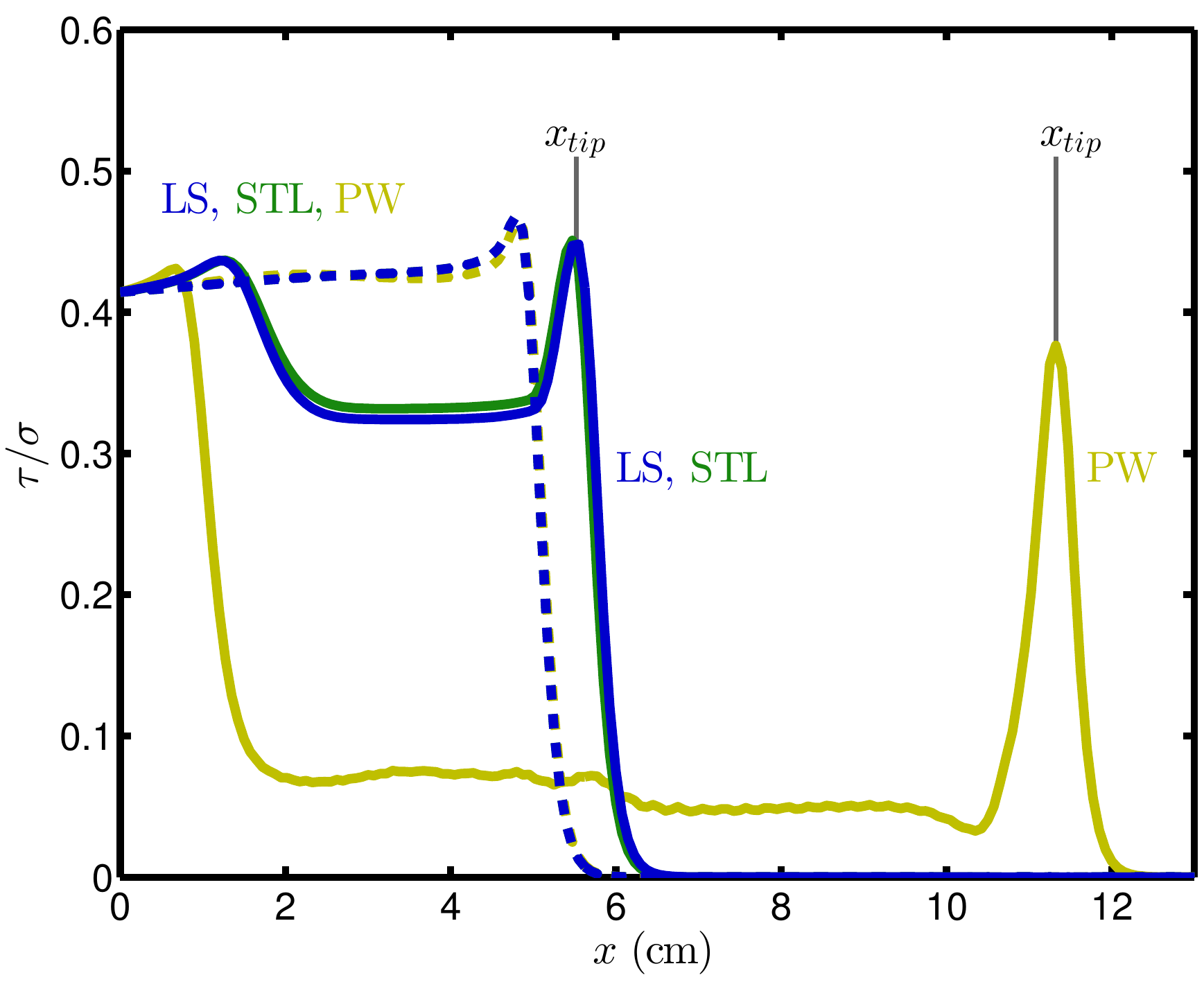}
\caption{The frictional stress 1s prior to the first event (dashed lines) and 1s after (solid lines).
The color code is as in Fig. \ref{fig:ss}. It is seen that the stress left at the tail of the rupture fronts,
$\tau_r$, is roughly homogeneous in space, and that it is much lower in the PW model than in the LS and STL models.
The location of the fronts after the event is marked by $x_{tip}$. The deeper penetration of the PW model, also shown in Fig. \ref{fig:fronts}, is clearly visible.}
\label{fig:before_after}
\end{figure}
A lot can be learned from the state of the interface after the rupture front has passed. In Fig.
\ref{fig:before_after} we plot the spatial distribution of the (normalized) friction stress just before the first rupture event
and immediately after it for the three variant models. In both of these states, the higher slip rates associated with the rupture fronts are not present (before the event they have not yet been generated and after the event they have died off), and the mechanical state is quasi-static. In line with the previous results, prior to the inception of the first event the
stress profiles in the three models essentially coincide. When the fronts propagate and eventually arrest, they leave behind them a residual stress
profile, which is much smaller in the PW model compared to the LS and STL models. This residual stress is approximately homogeneous in space and is lower than the stress prior to the event. The elastic energy release during this stress relaxation process is the driving force to frictional dissipation.

The approximate spatial homogeneity of $\tau$ left behind {\em any} rupture front when it arrests, allows us to estimate the loading force $f_d(t_{a})\!=\!\int_0^{L}\!\tau(x,t_{a})\,dx$ at the discrete arrest times $t_{a}$ (that is, there is $t_a$ corresponding to each rupture event). For that aim, we neglect the contribution to the integral in the region $x\!>\!x_{tip}(t_a)$, where $x_{tip}(t_a)$ is the location of the peak of $\tau$ slightly after a rupture front arrested (cf. Fig. \ref{fig:before_after}), and then assume that $\tau(x,t_a)$ can be replaced by a constant residual stress $\tau_r$, obtaining
\begin{equation}
\label{eq:approx_fd}
f_d(t_a)\!=\! \int_0^{L}\!\!\!\tau(x,t_a)\,dx \simeq\! \int_0^{x_{tip}(t_a)}\!\!\!\!\tau(x,t_a)\,dx \simeq  \tau_r \,x_{tip}(t_a) \ .
\end{equation}

To calculate $x_{tip}(t_a)$, we note that at the arrest times $t_a$ Eq. \eqref{eq:EOM1} takes the form $\tau_r\!\simeq\!\bar{G}H \pa_{xx}u(x,t_a)$ (i.e. in the range $0\!<\!x\!<\!x_{tip}(t_a)$ and neglecting inertia). With the approximate boundary conditions $u(x_{tip})\!\simeq\!\pa_x u(x_{tip})\!\simeq\!0$, this equation can be readily solved as
\begin{eqnarray}
x_{tip}(t_a)\simeq\sqrt{\frac{2\,\bar{G}\,H\,u(x\!=\!0,t_a)}{\tau_r}} \label{eq:xtip} \ .
\end{eqnarray}
This can be substituted in Eq. \eqref{eq:approx_fd} to give
\begin{eqnarray}
f_d(t_a)^2  \simeq 2\,\bar{G}\,H\,\tau_r\,v_d\,t_a \label{eq:fd_xtip} \ ,
\end{eqnarray}
where $u(x\!=\!0,t_a)\!=\!v_d\,t_a$ was used (which is, of course, valid at any time, not only at the discrete arrest times $t\!=\!t_a$).

The prediction in Eq. \eqref{eq:fd_xtip}, i.e. $f_d(t_a)^2\!\sim\!t_a$, is tested in the inset of Fig. \ref{fig:fd} for all three models over many events (i.e. this is a multiple-event property, not only a property of the first event, which was the focus of the discussion up to now). The analytic prediction is observed to be in favorable agreement with the numerical data for all three models, where the prefactor (slope) in the relation $f_d(t_a)^2\!\sim\!t_a$ is the same for the LS and STL models, but is significantly smaller for the PW model. These results show that $\tau_r$ is the same for every rupture event and lend direct support to the assumption that spatial variations of the residual stress left behind {\em any} rupture front can be neglected, consistent with the explicit stress profiles shown in Fig. \ref{fig:before_after} (for the first event in the three different models).

The latter observation allows us to extract $\tau_r$, the only unknown quantity in Eq. \eqref{eq:fd_xtip} (all other quantities are known parameters, which are the same for all three models), yielding $\tau_r/\sigma\!\simeq\!0.332$ for the LS and STL models and  $\tau_r/\sigma\!\simeq\!0.122$ for the PW model. The fact that the models that feature a nonmonotonic velocity dependence, i.e. the LS and STL models, give rise to an essentially identical residual stress $\tau_r$ is intimately related to the value of the steady state stress at the minimum of the friction curve (cf. Fig. \ref{fig:ss}), which is the same for both. Equation \eqref{eq:fd_xtip} then shows that the fact that the PW model produces a lower overall frictional resistance (and deeper force drops) compared to the LS and STL models is intimately related to the fact that the residual stress left behind the rupture fronts in the PW model is significantly lower than that of the LS and STL models. Furthermore, Eq. \eqref{eq:xtip} suggests an explanation for why the penetration depth, i.e. $x_{tip}(t_a)$, is significantly larger in the PW model than in the other two models.

The ``static friction coefficient'' $\mu_{static}$ is ordinarily defined as the tangential force, normalized by the normal force, needed to initiate global motion of the block. This force also corresponds to the peak of the loading curve. From this perspective, all of the spatiotemporal dynamics discussed up to now are precursory \cite{Rubinstein2004, Braun2009, Ben-David2010-fronts}, as they precede global motion which sets in only when a rupture front reaches the leading edge of the block (i.e. when $x_{tip}\!=\!L$). Hence, we can estimate $\mu_{static}$, which quantifies the global frictional resistance, as
\begin{equation}
\label{eq:static}
\mu_{static} \simeq \frac{f_d(x_{tip}\!\simeq\!L)}{\sigma L} \simeq \frac{\tau_r}{\sigma} \ ,
\end{equation}
where Eq. \eqref{eq:approx_fd} was used. This shows that the ``static'' frictional resistance of the interface, measured at slow loading velocities (here $v_d\!=\!10\mu$m/s), is influenced by dynamic processes at much higher slip rates and furthermore that the existence of velocity-strengthening friction behavior strongly affects $\mu_{static}$ through $\tau_r$ \cite{Ben-David2011, Capozza2012, Otsuki2013}.
\begin{figure*}
 \includegraphics[width=\textwidth]{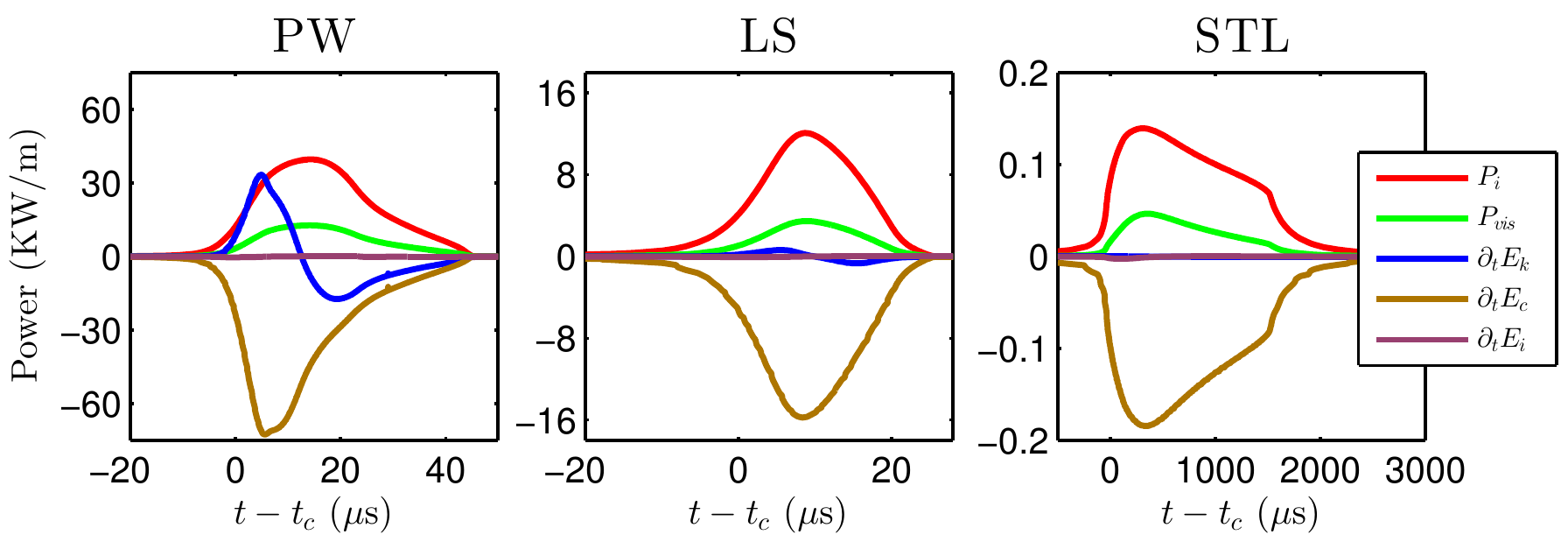}
 \caption{The rate of change of energies $\pa_t E_\gamma(t)\!=\!\pa_t\int_0^L\!\varepsilon\!_{_{\gamma}}\!(x,t)\,dx$ and dissipation rates $P_\gamma(t)\!=\!\int_0^L \!p_{_{\gamma}}\!(x,t)\,dx$ ($\varepsilon\!_{_{\gamma}}(x,t)$ and $p_{_{\gamma}}\!(x,t)$ are defined in Eqs. \eqref{eq:energies}) during the first event for the three models.}
 \label{fig:partition}
\end{figure*}

The results discussed above highlight two important points. First, an effectively constant residual stress $\tau_r$ is left behind rupture fronts in all of the models studied here. This property emerges spontaneously, unlike conventional slip-weakening models in which it is assumed a priori (see, for example, \cite{Ohnaka2000, Uenishi2003, Kammer2012} and the discussion in \cite{Cocco2002}). The value of $\tau_r$ depends on the existence of velocity-strengthening friction, which in turn has significant implications on the strength of the interface, as evident from Fig. \ref{fig:fd} and Eq. \eqref{eq:static}. Note also that the constancy of the residual stress $\tau_r$ implies that the mechanical fields associated with frictional shear cracks in 2D are well described by the classical theory of fracture \cite{Ilya2014}. Second, once $\tau_r$ is known, the arrest of rupture fronts is determined by global equilibrium conditions \cite{Taloni2014}, rather than by dynamic considerations (cf. Eq. \eqref{eq:approx_fd}).

\subsection{Energy partition: Dissipation and radiation}

As energy dissipation is at the heart of frictional phenomena, it will be interesting and instructive to consider the energy budget in
the system. As a starting point, we briefly remind the reader that the linear momentum conservation law of Eq. (\ref{eq:momentum}) can be transformed into a continuity equation for the energy density (using Hooke's law and integration by parts). The result reads
\begin{equation}
 \label{eq:continuity_3D}
 \pa_t\left(\tfrac{1}{2}\rho\left(\pa_t u_i\right)^2+\tfrac{1}{2}\epsilon_{ij}\sigma_{ij}\right)-\pa_j\Big(\sigma_{ij}\pa_t
u_i\Big)=0\ .
\end{equation}
The first term is the rate of variation of the energy density (both kinetic and elastic), and the second term is the divergence of the energy flux
vector. Their sum vanishes when energy is conserved.

Following the same procedure, one can derive the energy continuity equation for our model by combining Eqs. \eqref{eq:tauel} and \eqref{eq:EOM1}, obtaining
\begin{equation}
\label{continuity_1D}
\begin{split}
\pa_t\left(\varepsilon_k + \varepsilon_c + \varepsilon_i\right)&-\pa_x J=-p_i-p_{vis}\equiv -p \ ,
\end{split}
\end{equation}
where we defined
\begin{align}
&\varepsilon_k\equiv\tfrac{1}{2}\,\rho\, H\, (\pa_t u)^2\ , \qquad \varepsilon_c\equiv\tfrac{1}{2}\,\bar G\,H\,(\pa_x u)^2\ , \label{eq:energies}\\
  & \varepsilon_i\equiv \frac{\left({\tau^{el}}\right)^2}{2 G_0 A/\!h}\ , \qquad\quad\,\,\, J\equiv\bar G \,H\,v\, \pa_x u \ ,
\nonumber\\
  &p_i\equiv 2\,\varepsilon_i \frac{|v|}{D}\ , \qquad\qquad\,\,\, p_{vis} \equiv \tau^{vis}\,v\ . \nonumber
\end{align}
Here $\varepsilon_k$ is the kinetic energy density, $\varepsilon_c$ is the (bulk) linear elastic strain energy density, $\varepsilon_i$ is the interfacial elastic energy density and $J$ is the energy flux. The interfacial energy density, $\varepsilon_i$, is dissipated during sliding due to the rupture of asperities, resulting in a dissipation rate $p_i$ \cite{Note1}, in addition to the standard dissipation rate
$p_{vis}\!=\!\tau^{vis}\,v$.

Equation \eqref{continuity_1D} has the same structure as Eq. \eqref{eq:continuity_3D}, except for the non-vanishing dissipation rate $p$, which
exists because frictional dynamics are dissipative, and the existence of an interfacial elastic contribution $\varepsilon_i$ (both in the stored energy and in the dissipation power $p_i$).

\begin{table*}[th]
\begin{tabularx}{\linewidth}{|l|Y|Y|Y|}
\cline{2-4}
\multicolumn{1}{c|}{} & \bf{PW}& \bf{LS}& \bf{STL}  \\\hline
Velocity strengthening & Absent & Logarithmic & Linear \\\hline
Rupture propagation speed & $1540$ m/s& $166$ m/s& $\sim3$ m/s \\ \hline
Event's duration $\Delta t$ & $\sim\! 50\ \mu$s  & $\sim\! 35\ \mu$s  & $\sim\! 2000\ \mu$s  \\\hline
Total dissipated energy\footnote{Approximately equals to the bulk elastic energy released during the event, $\int_{\!\!_{\Delta t}}\!\!\!\!\pa_t E_c dt$.} & 1.4 J/m& 0.26 J/m  & 0.26 J/m\\\hline
Maximal dissipation rate\footnote{The maximum of $P\!=\!P_i\!+\!P_{vis}$.} & 52 kW/m & 16 kW/m & 0.19 kW/m \\\hline
Total radiated energy\footnote{The maximum of $E_k$.} & 0.27 J/m & 4.5 mJ/m  & 0.54 $\mu$J/m \\\hline
Penetration length $x_{tip}(t_a)$ & 11.3 cm& 5.52 cm & 5.46 cm \\\hline
\end{tabularx}
\caption{Summary of the main characteristics of the first rupture event in the LS, STL and PW models.}
\label{tab:dissipation}
\end{table*}
The quantities defined in Eq. \eqref{eq:energies} are densities that exhibit complex spatiotemporal behaviors during frictional instabilities (which result in rupture events). In order to gain some insight into these complex energy-exchange processes, it will be useful to consider the corresponding space-integrated quantities $E_\gamma(t)\!=\!\int_0^L\!\varepsilon\!_{_{\gamma}}\!(x,t)\,dx$ and $P_\gamma(t)\!=\!\int_0^L \!p_{_{\gamma}}\!(x,t)\,dx$.

The interplay between these various quantities during frictional instabilities (``events''), shown for all three models in Fig. \ref{fig:partition}, is an essential feature of interfacial dynamics. Our goal is to quantify generic energy-exchange processes during frictional instabilities \cite{Shi2008} and in particular to understand the differences between the three models in this respect. As the dynamics during frictional instabilities are much faster than typical loading rates, we expect them to be exclusively driven by the already stored elastic energy. That is, we expect the rate of change of the sum of bulk and interfacial elastic energies, $\pa_t(E_c\!+\!E_i)$, to be negative during an event. Figure \ref{fig:partition} clearly demonstrates this, and that $\pa_t E_i$ is negligible compared to $\pa_t E_c$ (hence we neglect the former compared to the latter in what follows).

The time integral of $\pa_t E_c$ over the event duration is the total energy released, which is a natural measure of the magnitude of the event (other measures exist as well). The elastic energy released is either being dissipated directly or is being first transformed into kinetic energy (``radiation''). Eventually, the kinetic energy is also dissipated. This generic picture is demonstrated in Fig. \ref{fig:partition} for all three models. In particular, it is observed that the dissipation contributions $P_i$ and $P_{vis}$ are comparable, where the former is typically larger than the latter. Kinetic energy generation (``radiation''), $\pa_t E_k\!>\!0$, is observed in the first part of the event. In the second part of the event $\pa_t E_k\!<\!0$, when the kinetic energy decays and is being dissipated.

While this generic qualitative picture is similar in all three models, there are large quantitative differences that we wish to discuss now. The main characteristics of the first rupture event in the LS, STL and PW models are summarized in Table \ref{tab:dissipation}. As we already know from Fig. \ref{fig:fronts}, the events are mediated by rupture fronts of vastly different velocities in the three models ($\sim\!10^3$m/s in the PW model, $\sim\!10^2$m/s in the LS model and $\sim$1m/s in the STL model). The event duration is about 40\% larger in the PW model as compared to the LS model, both in the few $10\mu$s range, while it is two orders of magnitude larger in the STL model ($\sim$ms). Despite the vast differences in the rupture propagation velocity and event duration, the total dissipated energy (which equals the amount of elastic energy released during the event) in the LS and STL models is essentially identical. This is in line with Fig. \ref{fig:fd}, which shows that the two models feature nearly identical stress drops and frictional resistance, and with Fig. \ref{fig:before_after} and the inset of Fig. \ref{fig:fd}, which show that the residual stress $\tau_r$ in the two models is essentially identical. This result clearly demonstrates that depending on the form of the velocity-strengthening friction branch (e.g. logarithmic vs. linear) one can observe events of the same magnitude (i.e. integrated dissipation/energy release) accompanied by very different dissipation rates (see Table \ref{tab:dissipation}). This result might be related to geophysical observations indicating that slow rupture does not necessarily imply smaller integrated slip and energy release \cite[especially Figure 5]{Peng2010}.

The total dissipation in the PW model is about $5.4$ times larger than the total dissipation in the LS and STL models, consistent with the much larger stress drops and the significantly reduced interfacial resistance observed in Fig. \ref{fig:fd}. Moreover, the amount of kinetic energy generated during the event is much larger in the PW model as compared to the other two models, and is about 19\% of the total energy released (though eventually it is also dissipated). In systems of larger heights $H$, this radiated kinetic energy will decay on longer timescales, allowing it to interact with remote boundaries. The kinetic energy generated in the STL model is negligibly small, while in the LS it makes about 1.7\% of the released energy (a similar value was reported in \cite{Shi2008}, although direct comparison is precarious). All in all, these results provide strong evidence that the existence and form of velocity-strengthening friction has significant implications on frictional dynamics and strength.

\section{Conclusions}

In conclusion, by studying the spatiotemporal dynamics in three variants of a realistic rate-and-state friction model under quasi-static side-loading conditions, we showed that the existence and form of velocity-strengthening friction may significantly affect various aspects of the frictional response of interfaces. These include the propagation velocity of coherent fronts that mediate interfacial rupture events, the emergence of slow rupture, the elastic energy released during events (i.e. their magnitude), the dissipation and radiation rates, and the global frictional resistance (strength). The clear connection between the existence of velocity-strengthening friction and slow rupture appears to be directly related to the recent experimental results of \cite{Kaproth2013}. It is also shown that events of similar magnitude (and hence stress drops) can be accompanied by substantially different dissipation and kinetic energy radiation rates.

Our theoretical results, together with extensive experimental evidence \cite{Bar-Sinai2013jgr}, highlight the need to quantitatively characterize the velocity-strengthening frictional response of interfaces, both experimentally and theoretically, and to systematically incorporate it into friction theory. Since frictional instabilities spontaneously lead to accelerated slip that probes relatively high-velocity properties of frictional interfaces, the latter -- which include velocity-strengthening friction -- affect the frictional response even under quasi-static loading conditions. This understanding may offer new ways to interpret existing observations in a broad range of frictional systems and to develop predictive theories of the dynamics of spatially extended frictional interfaces.


\end{document}